\documentclass[twocolumn,showpacs,preprintnumbers,amsmath,amssymb,prl]{revtex4}
\usepackage{graphicx}% Include figure files
\usepackage{dcolumn}% Align table columns on decimal point
\usepackage{bm}% bold math
\usepackage[usenames,dvipsnames]{color}% use as \textcolor{red}{******} 
\usepackage{ulem}

\begin{document}

\title{Double Exchange Ferromagnetism in the Peierls Insulator State}

\author{S. Nishimoto$^1$ and Y. Ohta$^2$}
\affiliation{$^1$Institut f\"ur Theoretische Festk\"orperphysik, IFW Dresden, 01171 Dresden, Germany}
\affiliation{$^2$Department of Physics, Chiba University, Chiba 263-8522, Japan}

\date{\today}

\begin{abstract}
We study the effects of opening of the band gap on the double exchange 
ferromagnetism.  Applying the density-matrix renormalization group method 
and an analytical expansion from the dimer limit to the one-dimensional 
double exchange model, we demonstrate for a relevant region of the exchange 
coupling that, in the weak dimerization regime, the Peierls gap opens in 
the fully spin-polarized conduction band without affecting its ferromagnetism, 
whereas in the strong dimerization regime, the ferromagnetism is destroyed 
and the Mott gap opens instead, leading the system to the antiferromagnetic 
quasi-long-range order.  An insulator version of the double exchange 
ferromagnetism is thus established.  
\end{abstract}


\maketitle

%
%\section{Introduction}
%

Magnetism and electronic transport properties of materials are 
closely related to each other, e.g., insulating transition-metal 
oxides are typically antiferromagnetic and ferromagnetism 
usually goes hand in hand with metallicity \cite{khomskii}, 
and elucidation of the mechanisms of this relationship is one of 
the major issues in the field of physics of strong electron 
correlations.  A well-known example is the double exchange 
ferromagnetism that occurs in mixed systems of localized spins 
and itinerant electrons interacting via the Hund's rule coupling, 
where the coherent motion of the itinerant electrons aligns the 
localized spins ferromagnetically to gain in kinetic energy of 
the itinerant electrons \cite{zener,anderson,degennes}.  

A question then arises as to what happens in this ferromagnetism 
if the coherent motion of electrons ceases due, e.g., to the 
opening of the band gap.  This issue comes into a real question 
in ferromagnetic hollandite K$_2$Cr$_8$O$_{16}$ \cite{hasegawa}, 
where the double exchange mechanism induces the three-dimensional 
(3D) full spin polarization in the system \cite{sakamaki}, and 
then the metal-insulator transition follows in its fully 
spin-polarized quasi-one-dimensional (1D) conduction band by the 
Peierls mechanism \cite{toriyama,nakao}, without affecting its 
3D ferromagnetism.  Thus, the uncommon ferromagnetic insulating 
(FI) state is realized in this material \cite{toriyama}.  

A naive answer to the above frequently asked question may then be 
that the ferromagnetism can survive if the band gap is small enough 
in comparison with the width of the conduction band and therefore 
the motion of conduction electrons, though not coherent, is not 
significantly suppressed.  However, to the best of our knowledge, 
no quantitative theoretical studies have been made on this issue 
whether this is actually the case.  

In this paper, we address this issue from the theoretical standpoint.  
We apply the numerical density-matrix renormalization group (DMRG) 
technique \cite{dmrg} and analytical expansion from the dimer limit 
to the 1D double exchange model that simulates the quasi-1D chain 
of K$_2$Cr$_8$O$_{16}$ and study the effects of opening of the band 
gap on the double exchange ferromagnetism.  We calculate the 
total-spin quantum number and charge gap of the system and extract 
the ground-state phase diagram of the model.  

We will thereby demonstrate for a relevant region of the exchange 
coupling that, in the weak dimerization regime, the Peierls gap 
opens in the fully spin-polarized conduction band without affecting 
its ferromagnetism, whereas in the strong dimerization regime, the 
ferromagnetism is destroyed and the Mott gap opens instead, due to 
the effective ``on-dimer'' Coulomb interaction, which leads the 
system to the antiferromagnetic quasi-long-range order.  
The metallicity itself is not therefore a necessary condition for 
the realization of the double exchange ferromagnetism and thus 
an ``insulator version'' of the double exchange ferromagnetism is 
established.  This is a route to realization of insulating 
ferromagnets, different from those discussed in doped LaMnO$_3$ 
where the orbital ordering plays an essential role 
\cite{hwang,mizokawa,geck}.  

%%%% FIG.1 %%%%
\begin{figure}[thb]
\begin{center}
\includegraphics[scale=0.58]{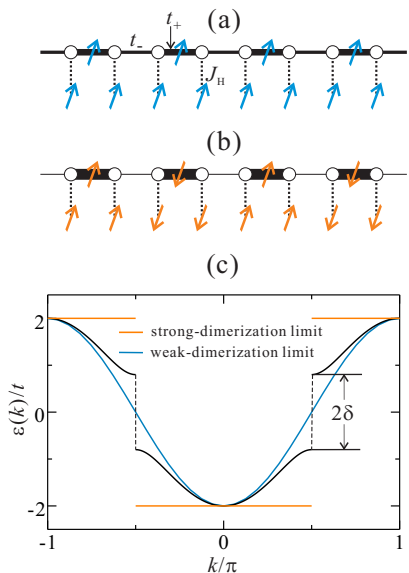}
\caption{(Color online) Schematic representations of the 
double exchange model in (a) the weak dimerization regime and 
(b) strong dimerization regime.  In (c), we illustrate the 
noninteracting band structure $\varepsilon(k)$ of our model with 
the lattice dimerization $\delta$ in the unfolded Brillouin zone.}
\label{fig1}
\end{center}
\end{figure}
%%%%%%%%%%%%%%%

%
%\section{Model and method}
%

Our model (see Fig.~\ref{fig1}) contains the terms of Peierls 
dimerization and Hund's rule coupling and is defined by the 
Hamiltonian
\begin{eqnarray}
&&{\cal H}=-\sum_{i=1,\sigma}^Lt_{i,i+1}
(c_{i,\sigma}^\dagger c_{i+1,\sigma}+{\rm H.c.})
-J_{\rm H}\sum_{i=1}^L{\bm s}_i\cdot{\bm S}_i\cr
&&{\bm s}_i=\frac{1}{2}\sum_{\sigma,\sigma'}
c_{i,\sigma}^\dagger{\bm\tau}_{\sigma,\sigma'}c_{i,\sigma'} ,
\label{ham}
\end{eqnarray}
where $c_{i,\sigma}^\dagger$ is the creation operator 
of an electron with spin $\sigma$ $(=\uparrow,\downarrow)$ 
at site $i$, ${\bm s}_i$ is the spin operator of a conduction 
electron at site $i$, ${\bm\tau}$ is the vector of Pauli 
matrices, and ${\bm S}_i$ is the quantum spin operator (of 
spin $1/2$) of a localized electron at site $i$.  
The hopping parameter between the nearest-neighbor sites is 
defined as $t_{i,i+1}=\big[1-(-1)^i\delta/2\big]t$
with the dimerization strength $\delta$ of $0\le\delta<2$; 
we in particular define $t_\pm=(1\pm\delta/2)t$.  
$J_{\rm H}$ is the strength of the Hund's rule coupling.  
$L$ is the number of sites in the system, where the site 
contains a conduction orbital and a localized spin.  
We confine ourselves to the case at quarter filling of 
conduction electrons $n=N/L=1/2$, where $N$ is the number 
of conduction electrons in the system.  
We introduce $\delta$ as a controllable input parameter 
rather than the order parameter of the spontaneous lattice 
dimerization since our purpose is to study the effects of 
$\delta$ on the electronic states of the model.  

We use the DMRG method to investigate the ground-state properties 
of the system Eq.~(\ref{ham}), where the open-end boundary conditions 
are applied.  We study the model with several lengths of $L=8$ to 
$20$ with keeping $m=1200$ to $3200$ density-matrix eigenstates 
in the renormalization procedure; in this way, the largest 
truncation error, or the discarded weight, is of the order of 
$10^{-11}$.  Note that we have to keep relatively-numerous 
density-matrix eigenstates to extract the true ground state from 
a number of nearly degenerate magnetic states, in particular, 
in the vicinity of the phase boundaries.  
The extrapolation to the thermodynamic limit $L\rightarrow\infty$ 
is made in the results presented unless otherwise indicated.  

%%%% FIG.2 %%%%
\begin{figure}[th]
\begin{center}
\includegraphics[scale=0.55]{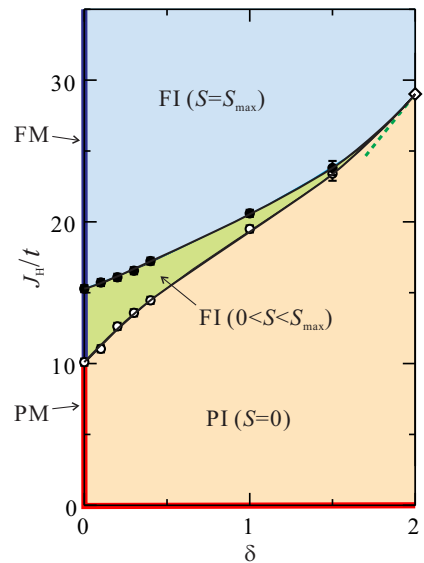}
\caption{(Color online) Calculated ground-state phase diagram 
of the 1D double exchange model at quarter filling with the 
lattice dimerization $\delta$.  We find the 
FI (ferromagnetic insulating), 
PI (paramagnetic insulating), 
FM (ferromagnetic metallic), and 
PM (paramagnetic metallic) phases.  
The FI-PI phase boundary at $\delta\rightarrow 2$ is determined 
by the analytical expansion from the dimer limit and is shown by 
the open diamond and dashed line.  
The FM and PM phases at $\delta=0$ or $J_{\rm H}=0$ are also 
indicated by the thick lines.  
}\label{fig2}
\end{center}
\end{figure}
%%%%%%%%%%%%%%%

%
%\section{Results of calculation}
%

\begin{figure}[htb]
\begin{center}
\includegraphics[scale=0.45]{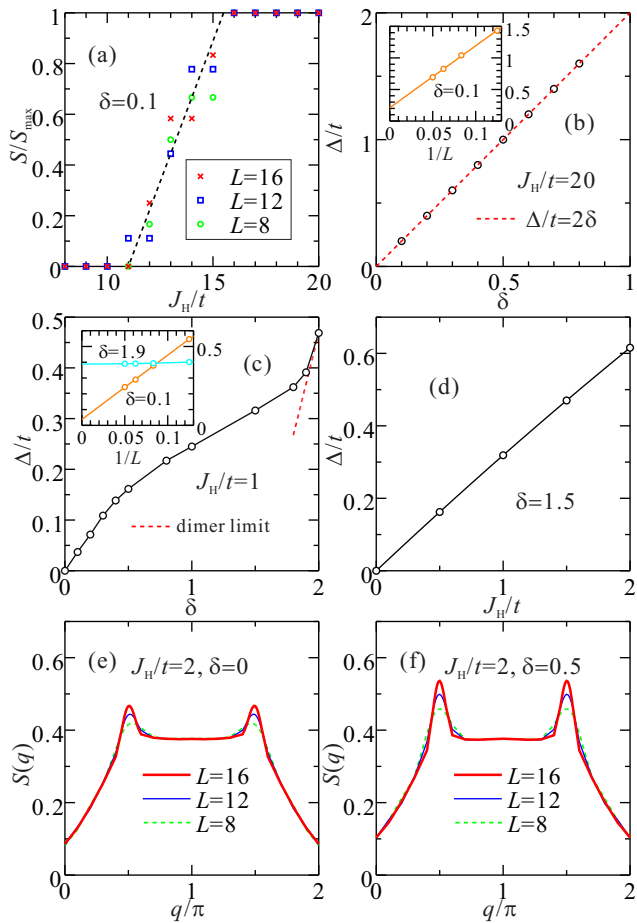}
\caption{(Color online) (a) Calculated normalized total-spin 
quantum number $S/S_{\rm max}$ as a function of 
$J_{\rm H}/t$.  Also shown are the calculated results for 
the charge gap $\Delta$: 
(b) $\delta$ dependence in the FI phase, 
(c) $\delta$ dependence in the PI phase (where the dotted line 
indicates the result of the strong-dimerization expansion), and 
(d) $J_{\rm H}/t$ dependence in the PI phase.  
Inset of (b) and (c) shows examples of the finite-size scaling 
of the gap.  
In (e) and (f), the calculated spin structure factors $S(q)$ 
in the PI phase without and with the lattice dimerization, 
respectively, are shown.  
}\label{fig3}
\end{center}
\end{figure}

First, let us present the total-spin quantum number $S$ of 
the system, which is calculated directly from the ground-state 
expectation value of ${\bm S}^2$ defined as 
$\langle{\bm S}^2\rangle=S(S+1)=
\sum_{ij}\big(\langle{\bm s}_i\cdot{\bm s}_j\rangle+
2\langle{\bm s}_i\cdot{\bm S}_j\rangle+
\langle{\bm S}_i\cdot{\bm S}_j\rangle\big)$.  
The results are given in Fig.~\ref{fig2} as a ground-state 
phase diagram.  We find that there are 
three phases, $S=0$, $0<S<S_{\rm max}$, and $S=S_{\rm max}$, 
depending on the values of $J_{\rm H}$ and $\delta$, 
where $S_{\rm max}$ denotes the full spin polarization.  
As expected, and in agreement with the previous calculations 
at $\delta=0$ \cite{kienert,henning,garcia,gulacsi}, 
the $S=S_{\rm max}$ phase appears when $J_{\rm H}$ is 
large and $S=0$ phase appears when $J_{\rm H}$ is small.  
And in between, the phase with the intermediate spin polarization 
also appears as in Ref.~\onlinecite{garcia} at $\delta=0$ 
(see also Fig.~\ref{fig3}(a) below).  
These phases are retained even when the dimerization 
$\delta$ is introduced.  We note that the two critical values 
of $J_{\rm H}$ that separate between the $S=S_{\rm max}$ 
and $S=0$ phases increase with increasing $\delta$ and 
that the region with the intermediate spin polarization 
becomes narrower and vanishes at $\delta\rightarrow 2$ 
(see below).  

Next, let us calculate the charge gap $\Delta$ defined as 
$\Delta=\lim_{L\rightarrow\infty}\Delta(L)$ with 
$\Delta(L)=\frac{1}{2}\big[E_0^{N+2}(L)+E_0^{N-2}(L)-2E_0^N(L)\big]$,
where $E_0^N(L)$ is the ground-state energy of the system 
of size $L$ with $N$ electrons.  The gap $\Delta$ is defined 
with the prefactor $1/2$, so that the single-particle band 
gap is identical to $\Delta$ in the present case where pairing 
interactions are absent.  
The results are shown in Figs.~\ref{fig3} (b)-(d).  
We find that the charge gap opens in the entire parameter 
space except at the lines $\delta=0$ and $J_{\rm H}=0$.  
The model at $J_{\rm H}=0$ is trivial, where the conduction 
electrons, decoupled completely from the localized spins, 
behave just as the noninteracting electrons, resulting in 
the PM phase.  

The model Eq.~(\ref{ham}) at $\delta=0$ on the other hand is 
highly nontrivial and much work has been done in recent 
years \cite{kienert,henning,garcia,gulacsi}: 
basically, there appears the ferromagnetic metallic (FM) phase 
for the large $J_{\rm H}$ region, which changes into the 
paramagnetic metallic (PM) phase when $J_{\rm H}$ becomes small.  
In addition, it has been claimed \cite{garcia,yunoki,dagotto} 
that there appears the region of phase separation in particular 
near half filling and the ``spiral'' phase with a long-wavelength 
antiferromagnetic correlations at the FM-PM phase boundary.  
Although we have not detected any indications of the phase 
separation at least at quarter filling in our accurate DMRG 
calculations with very large $m$ values, our results obtained 
are consistent with the results of the previous work 
\cite{kienert,henning,garcia,gulacsi}: we find that either 
the FM phase (when $J_{\rm H}$ is large) or PM phase 
(when $J_{\rm H}$ is small) is realized, and in between 
there is the partially spin-polarized metallic phase 
(see Figs.~\ref{fig2} and \ref{fig3}(a)) that may correspond 
to the spiral phase predicted in Ref.~\onlinecite{garcia}.  

%%%% FIG.3 %%%%
%%%%%%%%%%%%%%%

Let us then introduce the lattice dimerization $\delta>0$.  
The results are the following:  
In the $S=S_{\rm max}$ region, we find that the charge gap 
of $\Delta/t=2\delta$ opens as shown in Fig.~\ref{fig3}(b).  
This can simply be understood because in this region we 
have the noninteracting band of spinless fermions at half 
filling, and therefore the lattice dimerization $\delta$ 
opens the band gap of the size $\Delta/t=2\delta$.  The FI 
phase is thus realized.  Since the FM phase due to the 
double exchange mechanism is continuous to this FI phase, 
we may naturally refer to it as the insulator version 
of the double exchange ferromagnetism.  

In the $S=0$ region, we find that the gap actually opens as 
$\Delta\propto\delta$ in the weak dimerization limit as shown 
in Fig.~\ref{fig3}(c).  The size of the gap increases as 
the value of $J_{\rm H}$ increases as shown in 
Fig.~\ref{fig3}(d).  This phase with $\Delta>0$ may then 
be denoted as the paramagnetic insulating (PI) phase.  The 
Fourier transform of the spin-spin correlation function for 
the conduction electrons 
$S(q)=\frac{1}{L}\sum_{i,j}e^{iq(R_j-R_i)}
\langle{\bm s}_i\cdot{\bm s}_j\rangle$ 
(as well as that for the localized electrons, see Fig.~2 of 
Ref.~\onlinecite{garcia}) may characterize this phase.  
The calculated results are shown in Figs.~\ref{fig3} (e) and (f), 
where we find that the antiferromagnetic spin correlation of 
the wavevector of $q=\pi/2$ is enhanced and that the lattice 
dimerization inducing the localization of conduction electrons 
further enhances this correlation.  This results may therefore 
be interpreted as an enhancement by the lattice dimerization 
of the ``island'' state predicted in Ref.~\onlinecite{garcia}, 
where the high-spin ($S=3/2$) clusters, formed by a conduction 
electron coupled ferromagnetically with the two neighboring 
localized spins, are arranged antiferromagnetically with the 
quasi-long-range order \cite{garcia,fazekas}.  
In higher spatial dimensions, this phase may well fall into 
the true long-range antiferromagnetic order, resulting in the 
antiferromagnetic insulating (AFI) phase.  
This situation resembles that of the ``dimer-Mott'' phase 
\cite{penc,seo} in the dimerized Hubbard model at quarter 
filling although in the latter the spins are of $S=1/2$.  

In the $0<S<S_{\rm max}$ region, the charge gap also opens at 
$\delta>0$, where its size increases rapidly with increasing 
$J_{\rm H}/t$.  We thus have the FI phase here as well, which 
may be the insulating spiral phase continuous to the metallic 
one predicted in Ref.~\onlinecite{garcia}.  

Now, let us discuss the strong dimerization limit, where we 
start with the highly correlated clusters ${\cal C}_l$ 
($l=1,\cdots,L/2$) coupled weakly to each other through the 
hopping parameter $t_-$ (see Fig.~\ref{fig1}(b)).  
Each of the clusters consists of the two conduction orbitals 
and two localized spins.  In the ground state, the cluster contains 
one conduction electron (and two localized spins) and the internal 
three spins are fully polarized.  The lowest energy of the single 
cluster is $e(1)=-J_{\rm H}/4-t_+$, where the conduction electron 
is in the bonding state of the two conduction orbitals.  
The eigenstates of the cluster with the 0 and 2 conduction electrons 
are also derived exactly and the lowest energies are found 
to be $e(0)=0$ and $e(2)=-\sqrt{J_{\rm H}^2+16t_+^2}\big/2$, 
respectively.  In the strong-dimerization limit of the PI phase, 
we can therefore map our system Eq.~(\ref{ham}) onto an effective 
single-band Hubbard model defined in terms of the bonding orbital 
on each dimer.  The Hamiltonian may be written as 
\begin{eqnarray}
{\cal H}_{\rm eff}=t_{\rm eff}\sum_{i=1,\sigma}^{L/2}
(b_{i,\sigma}^\dagger b_{i+1,\sigma}+{\rm H.c.})
+U_{\rm eff}\sum_{i=1}^{L/2}n^b_{i,\uparrow}n^b_{i,\downarrow}, 
\end{eqnarray}
with the creation operator of an electron on the bonding orbital 
$b_i^\dagger=(c_{2i-1,\sigma}^\dagger+c_{2i,\sigma}^\dagger)/\sqrt{2}$ 
and $n^b_{i,\sigma}=b_{i,\sigma}^\dagger b_{i,\sigma}$.  
We obtain the effective hopping integral $t_{\rm eff}=t_-/2$ and 
effective ``on-dimer"" Coulomb interaction $U_{\rm eff}=e(2)+e(0)-2e(1)$.  
In this mapping, the localized spins contribute only to $U_{\rm eff}$ 
and their degrees of freedom are not explicitly involved in the 
operator $b_{i,\sigma}^\dagger$.  
An analytical expression for the charge gap may thus be derived as 
\begin{eqnarray}
\Delta&=&U_{\rm eff}-2t_-\cr
&=&-\sqrt{J_{\rm H}^2+4t^2(2+\delta)^2}/2+2t\delta+J_{\rm H}/2, 
\end{eqnarray}
which is valid up to the first order of $2-\delta$.  The result 
thus obtained is shown in Fig.~\ref{fig3}(c) as a dotted line, 
where we find that the agreement with the DMRG result is very good.  
In the FI phase, on the other hand, the charge gap is always 
$\Delta/t=2\delta$ in its entire region, independent of $J_{\rm H}$.  
The two gaps are therefore discontinuous at the phase boundary.  

Next, we derive the effective exchange interaction $J_{\rm eff}$ 
%(of the effective spin-$1/2$ Heisenberg model) 
between conduction electrons on the neighboring clusters.  
We first take a direct product of the isolated clusters 
($\prod_{l=1}^{L/2} {\cal C}_l$) to be the unperturbed ground 
state of the system.  Then, taking into account all the processes 
up to the second order of the hopping $t_-$, we obtain the 
expression for $J_{\rm eff}$ as 
\begin{eqnarray}
J_{\rm eff}&=&t_-^2\Big[\frac{(J_{\rm H}^2+8t_+^2)}
{(J_{\rm H}+4t_+)(J_{\rm H}^2+16t_+^2)}
+\frac{1}{8(J_{\rm H}+2t_+)}\cr
&+&\frac{(\sqrt{J_{\rm H}^2+16t_+^2}+4t_+)^2(J_{\rm H}+4t_+)}
{16J_{\rm H}t_+(J_{\rm H}^2+16t_+^2)}-\frac{3}{16t_+}\Big] ,
\label{jeff}
\end{eqnarray}
where we should note that, if only the lowest intermediate state 
is taken into account in the second-order process, we obtain 
$J_{\rm eff}=4t_-^2/U_{\rm eff}$, which is always positive 
(or antiferromagnetic).  We thus find that, depending on the value 
of $J_{\rm eff}$ (either positive or negative), the ground state 
realized is either the PI phase or the FI phase.  The critical value 
of $J_{\rm H}$ at $\delta \to 2$ is found to be $29.004t$, which 
determines the FI-PI phase boundary at $\delta \to 2$.  
The region of the intermediate spin state does not appear here.  
The result for the phase boundary obtained from Eq.~(\ref{jeff}) 
is shown as a dashed line in Fig.~\ref{fig2}, where we find again 
that the agreement with our DMRG result is very good, reinforcing 
the validity of our phase diagram of Fig.~\ref{fig2}.  

%
%\section{Summary}
%

In summary, we studied the effects of opening of the band gap 
on the double exchange ferromagnetism.  We applied the DMRG 
technique and analytical expansion from the dimer limit to 
the 1D double exchange model at quarter filling with lattice 
dimerization and obtained the ground-state phase diagram.  
We found three phases: the FI phase with the Peierls gap 
and full spin polarization, the PI phase with the Mott gap 
and dominant antiferromagnetic spin correlations, and the 
FI phase with the partial spin polarization.  The results 
for $J_{\rm H}/t\agt 15$ demonstrated that, in the weak 
dimerization regime, the Peierls gap opens in the fully 
spin-polarized conduction band without affecting its 
ferromagnetism.  Therefore, the metallicity itself is not a 
necessary condition for the realization of the double exchange 
ferromagnetism.  The concept of the insulator version of the 
double exchange ferromagnetism was thus established.  
In the strong dimerization regime, on the other hand, the 
ferromagnetism is destroyed at $J_{\rm H}/t\alt 29$ and the 
Mott gap due to the effective on-dimer Coulomb interaction 
opens there with the antiferromagnetic quasi-long-range 
order in the system.  

The uncommon FI state realized in K$_2$Cr$_8$O$_{16}$ 
\cite{toriyama} then means that this material is in the weak 
dimerization regime with the Peierls gap and full spin 
polarization.  
A recent experiment suggests \cite{yamauchi} that, by applying 
high pressures of $\agt 2$ GPa, the FM phase is suppressed very 
rapidly, while the metal-insulator transition remains almost 
unchanged, leading to the transition from the PM phase to 
the PI or AFI phase by lowering temperature.  
It may then be quite interesting to point out that, if the 
applied pressure decreases the value of $J_{\rm H}/t$, this 
might correspond to the phase change in the quasi-1D chains 
from the FI to PI phase as in our phase diagram given in 
Fig.~\ref{fig2}, where the intermediate spin state, or the 
spiral state of Ref.~\onlinecite{garcia}, may also be 
predicted to appear under high pressure.  
We hope that our work presented here will stimulate further 
searches for new phenomena and materials with intriguing 
magnetic and transport properties derived from the interplay 
between the double exchange and Peierls/Mott mechanisms.  

\begin{acknowledgments}
Enlightening discussions with R. Eder, D. I. Khomskii, 
T. Konishi, K. Nakano, T. Toriyama, Y. Ueda, and T. Yamauchi 
are gratefully acknowledged.  
This work was supported in part by a Kakenhi Grant 
No.~22540363 of Japan.  A part of computations was carried 
out at the Research Center for Computational Science, 
Okazaki Research Facilities, Japan.  
\end{acknowledgments}

\end{document}